\input epsf.tex
\def\DESepsf(#1 width #2){\epsfxsize=#2 \epsfbox{#1}}
\documentclass{ws-procs9x6}
\def\NPB{{ Nucl. Phys.} B}

\def\PLB{{ Phys. Lett.}  B}
\def\PRL{ Phys. Rev. Lett.}

\def\ZPC{{ Z. Phys.} C}

\begin{document}

\title{ Dark Matter, Muon g - 2 And
Other Accelerator Constraints}

\author{ R. Arnowitt$^{1}$ and B. Dutta$^{1,2}$}

\address{$^{1}$Center For Theoretical Physics,
Department of Physics, Texas A\&M University, College
Station, TX 77843-4242, USA\\
$^{2}$Department of Physics, University of Regina, Regina
SK, S4S 0A2, Canada}
\maketitle
\abstracts{ We review the current status of the Brookhaven muon g - 2 experiment, and
it's effects on the SUSY parameter space when combined with dark matter
relic density bounds, $b\rightarrow s\gamma$ and Higgs mass constraints. If the 
3$\sigma$ deviation of g - 2 from the Standard Model value is correct, these
data constrain the mSUGRA parameter space strongly, i.e. 300 GeV 
$\stackrel{<}{\sim} m_{1/2}\stackrel{<}{\sim}$
850 GeV, and $m_0$ ( at fixed $\tan\beta$, $A_0$) is tightly constrained (except at
very large $\tan\beta$). Dark matter detection cross sections lie within the
range accessible to future planned experiments. A non-universal gluino soft
breaking mass however can greatly reduce the lower bound on $m_{1/2}$ (arising
from the $b\rightarrow s\gamma$ constraint) allowing for relatively light neutralinos,
while non-universal Higgs $H_2$ mass can lead to new regions of allowed relic
density where the detection cross sections can be increased by a factor of
10 or more.
}

\section{Introduction}

While current Tevatron and LEP measurements have not greatly constrained
the SUSY particle spectra, there are a number of quantities, which if
accurately measured and if accurate theoretical calculations existed, could
greatly limit the SUSY parameter space of a given model, and thus allow
significant predictions of what might be expected at the LHC and what might
occur in the next round of dark matter detector experiments. We consider
here the following quantities: the muon g - 2, the light Higgs mass $m_h$,
the $b \rightarrow s\gamma$ branching ratio and the amount of dark matter. If
accurately determined, these would greatly restrict the SUSY parameter
space for a variety of models. We will first examine these within the
framework of mSUGRA models (with R parity invariance) and then show some
non-universal models which could moderate somewhat the mSUGRA constraints.

\section{The Muon g - 2 Anomaly: The Saga Continues!}

We review the current situation with the muon g - 2 magnetic moment
anomaly. Recall that in 2001, the Brookhaven E821 experiment reported their
high precision measurement of   $a_\mu = (g - 2)/2$. Based on the then best
theoretical calculation of the Standard Model value \cite{dh}, they reported a
2.6$\sigma$ deviation. Unfortunately, a sign error was subsequently found in
the ``scattering of light by light" (LbL) diagrams, which reduced the effect
to only 1.6sigma. Since that time the following has happened:

(i) New E821 data (the $\mu^+$ 2000 data) has been analyzed reducing the
experimental error in $a_\mu$ by a factor of two\cite{BNL}. The current world average
is now\begin{equation}
                                a_\mu^{\rm exp} = 11 659 203 (8)\times10^{-10}
\end{equation}
i. e. a measurement at the level of 0.7ppm! One interesting feature of the
Brookhaven measurements is the stability of their central value (with the
error flag successively being reduced).

(ii) New data has come from Novosibirsk (CMD-2) and Beijing (BES) on $e^+ +
e^- \rightarrow $ hadrons, and from ALEPH and CLEO on tau decay into two and four
pions. These may be used to calculate the hadronic contributions to the SM
$a_\mu$ prediction to get a more accurate determination of any deviation
between theory and experiment that  may exist.

Two groups \cite{dav,hag} have now used this new data to reevaluate $a_\mu^{\rm SM}$, and we
briefly review their results. $a_\mu^{\rm SM}$ can be divided into the following parts:
\begin{equation}
               a_\mu^{\rm SM} = a_\mu^{\rm QED} + a_\mu^{\rm weak} + a_\mu^{\rm had}
\end{equation}
where
\begin{equation}      a_\mu^{\rm had} = a_\mu^{\rm LO} + a_\mu^{\rm LbL} + a_\mu^{\rm HO}
\end{equation}

The QED and weak contributions to $a_\mu$ are well established, and the higher
order (HO) hadronic contribution appears to be in good shape. With the
corrected sign, the light by light (LbL) contribution has been evaluated by
several groups \cite{knecht} with general agreement. We use here the value
$a_\mu^{\rm LbL} =
[8.6 \pm 3.5]\times10^{-10}$. Current difficulties arise from the leading order in
alpha (LO) hadronic contributions. This quantity can be calculated using a
dispersion relation:
\begin{equation}
  a_\mu^{\rm LO} ={\alpha^2(0)\over {3\pi^2}}\int^{\infty}_{4m^2_{\pi}}ds{K(s)\over s}R(s)
\end{equation}
where $K(s)$ is the QED kernel and
$R(s) = \sigma(e^+e^-\rightarrow{\rm hadrons})/\sigma(e^+e^-\rightarrow
\mu^+\mu^-)$. The integral is strongly weighted at low energy, i.e. about 90\%
comes from $\sqrt{s} < 1.8$GeV and about 75\% from $e^+e^- \rightarrow\pi \pi$ (from the
$\rho$).

Two procedures have been used to evaluate the dispersion integral. The
first method uses the $e^+ e^-$ cross sections of CMD-2, BES and a
large amount of earlier data to calculate R(s).  However, it should be
noted that the $\sigma(e^+e^- \rightarrow {\rm hadrons})$ to be used in Eq.(4) is a ``bare"
cross section, i.e. the experimental data must be corrected for initial
state radiation, photon vacuum polarization and electron vertex loop
effects (and not carrying this out correctly has led to some errors in past
analyses.) Carrying out this analysis, Ref.\cite{dav} finds a discrepancy between the
experiment and the SM of $\Delta a_\mu = 33.9(11.2)\times10^{-10}$ a 3.0$\sigma$
effect, while Ref.\cite{hag} finds $\Delta_\mu = 35.13(10.63)\times10^{-10}$ i.e a 3.3$\sigma$
effect. Thus the two analyses give consistent results.

The second method makes use of the tau decay data of ALEPH and CLEO into
2$\pi$ and 4$\pi$ final states. Using CVC, the isovector form factor can be used
to construct $\sigma$($e^+e^-\rightarrow 2\pi$,$4\pi$) for $s \stackrel{<}{\sim}3$GeV (which is most of the
important region). However, in this case, one must include corrections due
to the breaking of CVC. Major contributions to this come from the $\pi$ mass
differences and the short distance radiative corrections (treated by chiral
perturbation theory). Carrying out this analysis, Ref.\cite{dav} finds from the tau
data a discrepency of   $16.7(10.7)\times10^{-10}$ i. e. only a 1.6$\sigma$ effect.
Further, one cannot simply average the two proceedures of calulating
$a_\mu^{\rm LO}$. To point up the problem \cite{dav} reversed the proceedure, and using the
$e^+ e^-$ data plus the CVC breaking corrections, predicted the tau branching
ratios into 2$\pi$ and 4$\pi$ final ststes and then compared this to the
experimental ALEPH and CLEO data. They found that the prediction fails for
the $\pi^- \pi^0$ state by 4.2$\sigma$, and fails for the 2$\pi^- \pi^+ \pi^0$ state by
3.5$\sigma$. Thus the two approaches are statistically inconsistent with each
other.

At this point, there is no explanation for the disagreement between the two
types of analysis. The discrepancy appears to be too large to be attributed to
lack of understanding of the CVC breaking effects. Thus the question of
whether the muon magnetic moment anomaly implies new physics is once again
unclear. There will be more $e^+ e^-$ data from CMD-2, BES and also KLOE and
BABAR. The B-factories may also be able to measure the tau decays. Finally,
the Brookhaven E821 experiment has $3\times 10^9$ $\mu^-$ events it is currently
analyzing. (Results may be out by early next year.)  It also has future
plans for running the experiment further.
\begin{figure}\vspace{-0cm}
 \centerline{ \DESepsf(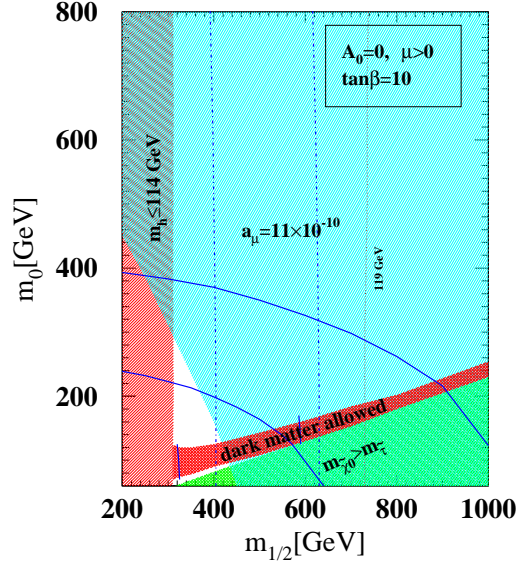 width 5.95 cm) } 
\caption {\label{fig1} mSUGRA allowed regions for $\tan\beta = 10$, $A_0 = 0$, $\mu >0$.
The dot-dash lines are the bounds for a Linear Collider(LC) $\tilde\chi^0_1
-\tilde\chi^0_2$ signal (for $\sqrt{s} = 500$ GeV and $800$ GeV), and the
curved lines are for the stau pair production signal.}
\end{figure}

\begin{figure}\vspace{-0cm}
 \centerline{ \DESepsf(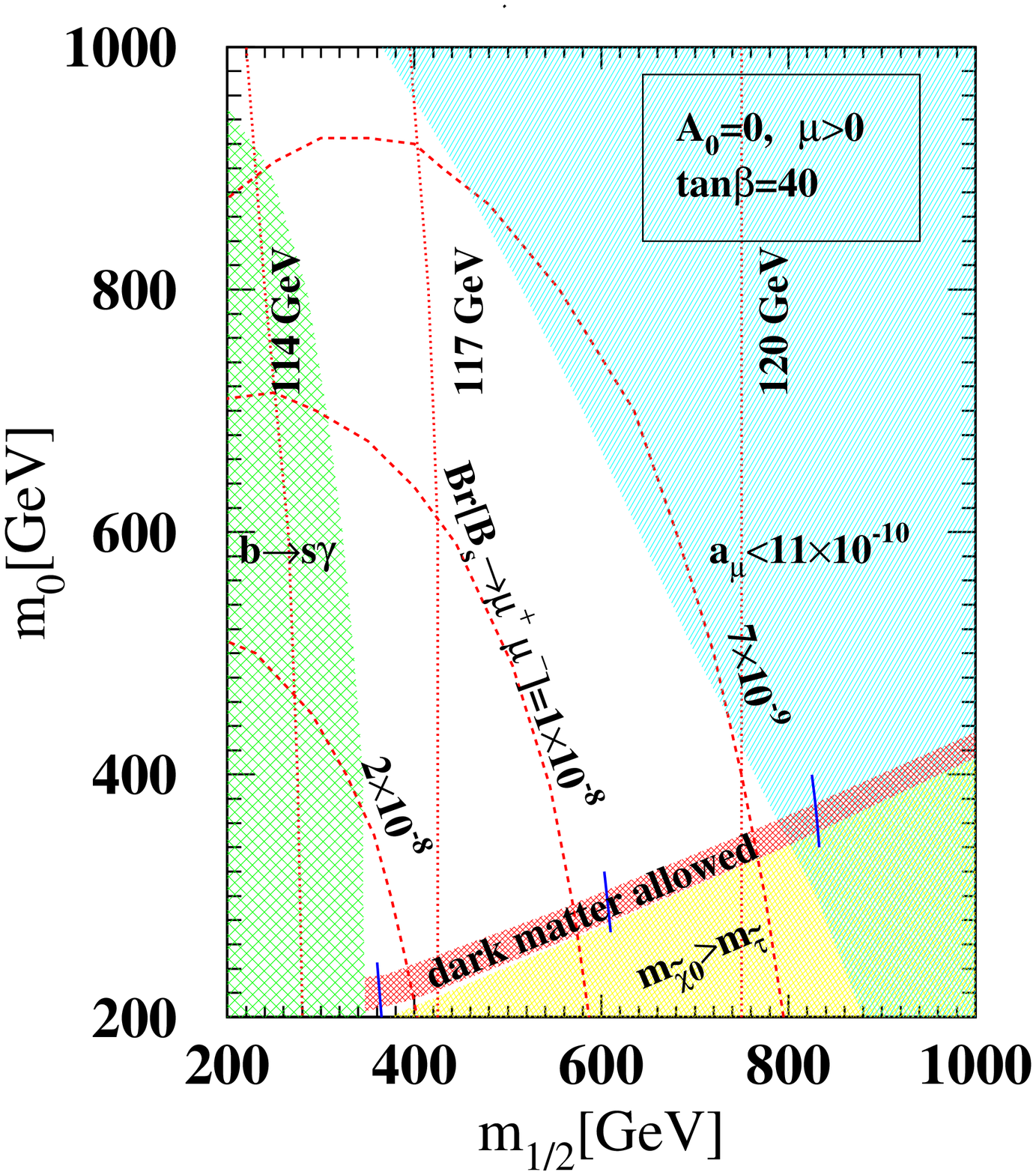 width 5.95 cm) } 
\caption {\label{fig2}  Same as Fig.1 for $\tan\beta  = 40$, $A_0 = 0$, $\mu >0$.The dashed
line give contours of $B_s \rightarrow \mu \mu$ branching ratios at the
Tevatron. The vertical dotted lines are Higgs mass contours. The short
vertical lines are $\sigma_{\chi^0_1-p} = 3\times10^{-8}$pb (lower) and
$1\times 10^{-9}$pb (upper).}
\end{figure} 
\section{mSUGRA Model}
In SUSY models, one generally has a contribution to $a_\mu$ in addition to
the SM piece \cite{g-2,g-22}. This arises from loops with chargino and sneutrino or
neutralino and smuon intermediate states, with a magnetic field attached to
any charged particle. In general these contributions are not small. If the
analysis leading to a 3$\sigma$ effect is valid, and if one attributes the
deviation to the SUSY correction, one gets a significant constraint on the
SUSY parameter space, with the SUSY spectrum required to be relatively low
and easily within the reach of the LHC. On the other hand, if the 1.6 sigma
analysis turns out to be correct implying only a small SUSY contribution is
possible, then squark and gluino mass spectrum would be pushed to the TeV
domain. Thus the resolution of the current ambiguity in $a_\mu^{\rm SM}$ is very
important. In this section, we analyze these matters within the framework
of the mSUGRA model with R parity invariance.

mSUGRA is the simplest SUGRA model in that it depends only on four
parameters and one sign: $m_0$ (the scalar soft breaking mass at $M_G$),
$m_{1/2}$ (the gaugino mass at $M_G$), $A_0$ (the cubic soft breaking
parameter at $M_G$), $\tan\beta = <H_2>/ <H_1>$ (at the electroweak scale),
and the sign of $\mu$ (the Higgs mixing parameter in the
superpotential:$\mu H_1 H_2$). We examine here the parameter range $m_0
>0$, $m_{1/2} < 1$TeV (which corresponds to the gluino mass bound accessible to
the LHC: $m_{\tilde g} < 2.5$GeV), $2 \leq\tan\beta \leq55$, and $|A_0|
\leq 4m_{1/2}$. One starts the analysis at $M_G$ and uses the
renormalization group equations (RGE) to go down to the electroweak scale.
Thus all SUSY masses and cross sections are determined in terms of these
four parameters and one sign. The details of carrying out this analysis
including all coannihilation effects in the relic density analysis can be
found in e. g. \cite{bdutta}.

The mSUGRA model allows one to calculate an array of quantities that can be
measured and hence can be used to restrict the parameter space. The include
the neutralino relic density, the $b \rightarrow s + gamma$ branching ratio
and the Higgs mass bound. For the relic density we take a $2\sigma$ range
around the current central value \cite{turner}:
\begin{equation} 0.07 \leq \Omega_X h^2 \leq 0.21
\end{equation}
For the $b \rightarrow s + \gamma$ decay (which has both systematic and
theoretical uncertainties) we take a relatively broad range around the CLEO
central value \cite{bsgamma} and use the LEP bound for $m_h$ \cite{higgs1}:
\begin{equation} 1.8\times10^{-4} \leq B(b \rightarrow s\gamma) \leq
4.5\times10^{-4}; \,\,\,m_h >114 {\rm GeV}
\end{equation}

In addition there is the LEP bound on the chargino mass of $m_{\chi_1^\pm}
> 103$GeV\cite{aleph}.
\begin{figure}\vspace{-0cm}
 \centerline{ \DESepsf(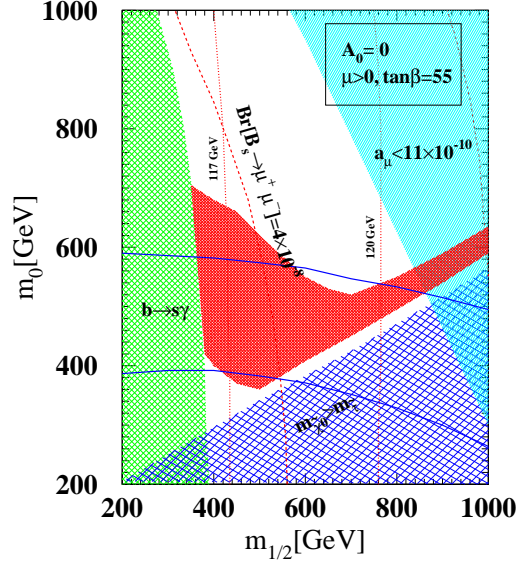 width 5.95 cm) } 
\caption {\label{fig3}Same as Fig. 2 for $\tan\beta = 55$, $A_0 = 0$, $\mu >0$ and $m_t =
175$GeV, $m_b = 4.25$.}
\end{figure}

\begin{figure}\vspace{-0cm}
 \centerline{ \DESepsf( 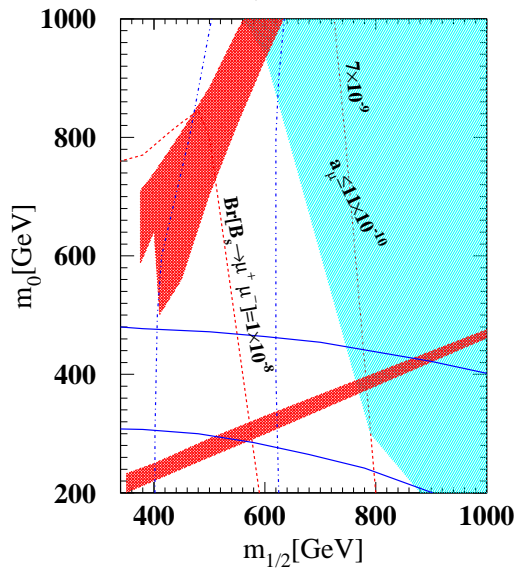 width 5.95 cm) } 
\caption {\label{fig4} Same as Fig. 2 for $\tan\beta = 40$, $A_0 = m_{1/2}$, $\mu
>0$ for non-universal soft breaking of the Higgs masses with $\delta_2 = 1$.
The lower shaded band is the usual allowed stau-neutralino coannihilation
band, and the upper band is the new region arising from the non-universal
Higgs masses due to increased annihilation through the $Z$-channel.}
\end{figure}
In spite of the large errors in the experimental data that still are
present, Eqs.(5,6) put significant constraints on the SUSY parameter space
in that they require that $m_{1/2}\stackrel{>}{\sim}(300-400)$ GeV across the full
parameter space. This is illustrated in Figs. (1-3) for $\tan\beta = 10$,
40, 55. If we assume the 3$\sigma$ deviation for the muon $a_\mu$ is valid,
then the 2$\sigma$ bound from the central value is $11\times10^{-10}$. For
low $\tan\beta$, e.g. $\tan\beta = 10$ (Fig. 1), the combined constraints
then leave very little available parameter space. For higher $\tan\beta$,
e.g. $\tan\beta = 40$ (Fig.2), more allowed region exits, but the
$\chi^\pm_1$ mass constraint combined with the $a_\mu$ constraint
eliminates all the ``focus point" region \cite{feng} of large $m_0$ and low
$m_{1/2}$. For very high $\tan\beta$, e.g. $\tan\beta = 55$ (Fig.3) a new
region opens up producing a ``bulge" at low $m_{1/2}$ due to rapid s-channel
annihilation of the $A$ and $H$ Higgs bosons (as their mass is
significantly reduced). Throughout the entire range, the $\chi^0_1$ -
proton cross sections for direct detection of Milky Way dark matter lie in
the range of $(9\times10^{-8} - 5\times10^{-10})$pb, a range that is
expected to be accessible to future large scale experiments.

\section{Non-Universal Models}

We consider here two types of non-universal soft breaking  at the GUT scale: for the gaugino
masses and for the Higgs masses. The first is of interest
in that it can soften significantly the lower bound produced by the
$b\rightarrow s\gamma$ and $m_h$ constraints. Thus if one assumes at
$M_{\rm GUT}$ that the gluino mass is $m_{1/2} ( 1 + \tilde\delta_3)$
(where $m_{1/2}$ is the universal mass), then for $\tilde\delta_3 =
1$, one finds e.g. for $\tan\beta = 50$,that the lower bound on $m_{1/2}$
is reduced to 185 GeV corresponding to a neutralino mass of 75 GeV. (The
constraint from $m_h$ is also softened, though the $a_\mu$ constraint
becomes somewhat stronger).

If the $H_2$ mass is increased at $M_G$, new effects also occur. Thus
writing $m_{H_2}^2 = m_0^2 (1 + \delta_2)$, where $m_0$ is the universal
scalar mass, then a new annihilation channel in the relic density analysis
can open up for low $m_{1/2}$ and high $m_0$ due to rapid annihilation
through the s-channel $Z$ pole. This is shown in Fig. 4 for $\tan\beta =
40$. In this new region, the neutralino-proton cross section also increases
by a factor of 10 or more, allowing cross sections in the range $(10^{-6} -
10^{-7})$pb.

\section{Conclusions}

The muon magnetic moment anomaly can produce strong constraints on the
allowed region of SUSY parameter space. Thus if the 3$\sigma$  deviation
analysis is correct, then a 2$\sigma$ bound from the central value gives an
upper bound of $m_{1/2}\stackrel{<}{\sim}850$GeV when combined with the relic density
constraint. The $b\rightarrow s\gamma$ and $m_h$ constraints in the mSUGRA
model, produce a lower bound of $m_{1/2} > (300 - 400)$ GeV, thus bounding
the parameter space in a region easily accessible to the LHC. Dark matter
detection cross sections then lie in a region accesssible to future dark
mater experiments.However, should the $a_\mu$ anomaly be smaller than $\sim
10\times10^{-10}$ the squark and gluino spectrum will be pushed into the
TeV domain. Non-universal models can modify these constraints. Thus an
increase in the gluino mass at $M_G$ can signficantly decrease the the lower
bound on $m_{1/2}$ (and hence on the neutralino mass), while an increase of
the $H_2$ mass at $M_G$  can give rise to new allowed regions of relic density
with detection cross sections increased by a factor of 10 or more.

 This work was supported in part by the National Science
Foundation Grant PHY - 0101015.

\end{document}